\documentclass[bibyear]{aa} 

\usepackage{graphicx}
\usepackage{txfonts}
\usepackage{color}

\begin{document}

   \title{Impact of young stellar components on quiescent galaxies: deconstructing cosmic chronometers}
   \subtitle{}

   \author{M. L\'opez-Corredoira\inst{1,2}, A. Vazdekis\inst{1,2}}

   \institute{$^1$ Instituto de Astrof\'\i sica de Canarias, 
E-38205 La Laguna, Tenerife, Spain\\
$^2$ Departamento de Astrof\'\i sica, Universidad de La Laguna,
E-38206 La Laguna, Tenerife, Spain }

   \date{Received xxxx; accepted xxxx}

 
  \abstract
  {Cosmic chronometers may be used to measure the age difference between passively evolving 
   galaxy populations to calculate the Hubble parameter $H(z)$ as a function of redshift $z$. 
   The age estimator emerges from the relationship between the amplitude of the rest frame
   Balmer break at 4000 \AA \ and the age of a galaxy, assuming that there is one single stellar population 
   within each galaxy.}
   {First, we analyze the effect on the age estimates from the possible contamination ($<2.4$\%  of the stellar mass in our high-redshift sample) of a young component of  $\lesssim 100$ Myr embedded within the predominantly old population of the quiescent
galaxy. Recent literature has
    shown this combination to be present in very massive passively evolving galaxies. Second, we evaluate how the 
    available data compare with the predictions of nine different cosmological models.}
   {For the first task, we calculated the average flux contamination due to a young 
    component in the Balmer break from the data of 20 galaxies at $z>2$ that included photometry 
    from the far-ultraviolet to near-infrared at rest. For the second task, we compared the data 
    with the predictions of each model, using a new approach of distinguishing between 
    systematic and statistical errors. In previous work with cosmic chronometers, these have 
    simply been added in quadrature. We also evaluated the effects of contamination by a young 
    stellar component.}
   {The ages inferred using cosmic chronometers represent a galaxy-wide average rather than
a characteristic of the oldest population alone. The average contribution from the young 
    component to the rest luminosity at 4000 \AA \ may constitute a third of the luminosity in some samples, which means that this is far from negligible. This ratio is significantly dependent on stellar mass,
proportional to $M^{-0.7}$. Consequently, the measurements 
    of the absolute value of the age or the differential age between different redshifts are 
    at least partially incorrect and make the calculation of $H(z)$ very inaccurate. Some
cosmological models, such as the Einstein-de Sitter model or quasi-steady state cosmology, which are rejected under the assumption of a
purely old population, can be made compatible with the predicted ages of the Universe as a function of redshift if we take this
contamination into account. However, the static Universe models are rejected by these $H(z)$ measurements, even when this contamination
is taken into account.}
   {}

   \keywords{Galaxies: evolution -- Galaxies: high-redshift -- Cosmology: observations}

\titlerunning{Cosmic chronometer criticism}
\authorrunning{L\'opez-Corredoira \& Vazdekis}

   \maketitle
%

\section{Introduction}

The geometry of the Universe and the reality of its expansion may be tested by means of
various analyses of the galaxy distribution, either in space or in time. The task is
not simple and straightforward because there are many selection effects that mix
truly cosmological effects with the evolution of galaxies, redshift distortions, and other 
influences, thus limiting the capability of the tests to constrain cosmological models 
without strong systematic effects (see, e.g., Baryshev \& Teerikorpi 2012; 
L\'opez-Corredoira 2017). 

For the time measurements, cosmological models provide precise predictions of the age 
of the Universe at a given distance, thus allowing a comparison with any chronometric 
measurements that establish how fast time changes with redshift. An example of an attempt 
to constrain cosmological models with this type of time measurement is the derivation of 
the Hubble parameter $H(z)$ as a function of redshift (Jim\'enez \& Loeb 2002; Moresco 
et al. 2012, 2016; Ratsimbazafy et al. 2017), using massive (with a stellar content 
$>10^{11}$ M$_\odot $), passively evolving galaxies; this method is called  ``cosmic chronometers''. 
This technique uses the age difference between passively evolving galaxies at different
redshifts to calculate the expansion rate of the Universe. In
addition to its use as a tool 
for constraining the cosmological parameters in the standard model, it has also been used 
to successfully fit the data with the predictions of other competing alternative models, 
such as the $R_h=ct$ Universe (Melia \& Maier 2013; Melia \& McClintock 2015; Wei et al. 2017; Leaf \& Melia 2017; Melia \& Yennapureddy 2018).

We here extend the way different cosmologies
are tested using this method in several important ways. We begin with a discussion of the misconceptions concerning this method, 
both in terms of the a priori assumption of a single stellar population (\S \ref{.young}), 
and in the handling of the statistical and systematic errors in the analysis 
(\S \ref{.stat}). We carry out a comparative analysis of nine different models in 
\S \ref{.fitting} and dedicate a final section (\S \ref{.conclusion}) to the conclusions 
and further discussion.

\section{Critical assessment of the $H(z)$ measurements}
\label{.young}

The main flaw of the cosmic chronometer method is that it assumes that early-type 
galaxies have only a single burst stellar population (SSP) or an extended stellar
star formation with a very fast decay (star formation ratio $\propto e^{-t/\tau }$ with $\tau <0.3$ Gyr), which is in practice comparable to a Dirac's delta in the distribution of ages
of an SSP, especially when observed in the nearby Universe. Observational evidence 
suggests that this assumption is not valid, given that one SSP cannot fully 
describe early-type galaxies in general at any redshift and any stellar mass (Burstein et al. 1988; Vazdekis et al. 1997, 2016; Yi et al. 2007, 2011; Atlee et al. 2009; Salim \& Rich 2010; Mok et al. 2014; L\'opez-Corredoira et al. 2017 [hereafter L17]),
and because there is residual gas even at high-redshift quiescent galaxies (Concas et al. 2017; Gobat et al. 2017). It is relevant that this young population for the most massive galaxies affects the samples used for cosmic chronometers.
Specifically, fits to the galaxy spectra based on one SSP do not provide consistent results for the blue and red wavelengths at high redshifts; at least two SSPs are necessary to match the photometric and spectroscopic data, with the bulk of the stellar population being very old and the remaining contribution coming from a young component that is due to residual star formation (Oser et al. 2010).
The evidence of this second young burst arises because the UV spectral range is almost
unnoticed in the visible.
This ``contamination'' represents a drawback with age estimator methods that are based on single lines or breaks, such as the Balmer break used in cosmic chronometers, thus (perhaps severely) weakening the application of this technique to constrain cosmological parameters.

\begin{figure}[htb]
\vspace{1cm}
\centering
\includegraphics[width=8cm]{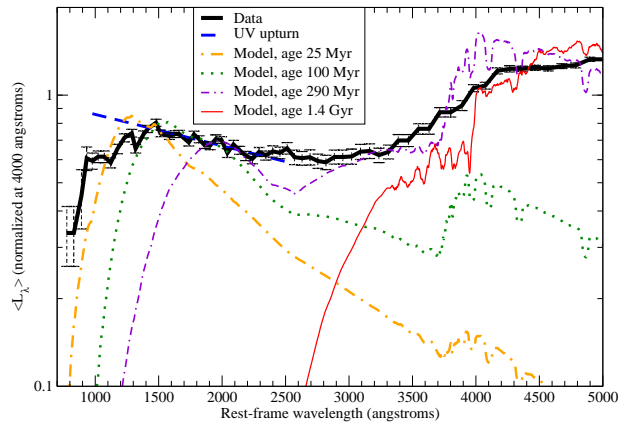}
\vspace{0.5cm}
\caption{Log-linear plot of the average normalized (such that $L_{4000\ \AA}=1$) luminosities of the 20 galaxies in the sample of L17. The dashed line represents the expectation for an UV upturn. The models are derived from the Bruzual \& Charlot (2003) synthesis model for different ages and metallicity $Z=0.05$, binned to $\Delta \lambda \sim 50$ \AA. For illustrative purposes the
normalization of the fluxes of the young population models have been varied artificially to match the UV upturn.}
\label{Fig:avfitphotl2}
\end{figure}

L17 demonstrated that some age-sensitive breaks, such as the 
Balmer break at 4\,000 \AA \ at rest, have contributions from significant fractions of light of two such components, therefore they cannot serve as a simple, direct indicator of the age  of the oldest population, but instead represent a luminosity-weighted age of the young and the old  populations. One might think that such double populations affect only a few quiescent galaxies and that most of them are reasonably well fit with only one SSP. However, observational evidence to support this is lacking; 
quite to the contrary, photometry in the wavelength of the far UV at rest is needed to characterize the young component, and 
this is not done in most of the quiescent galaxies at low-intermediate redshifts, 
which are typically observed only in the visible.
In the few cases where UV 
photometry at rest is available, a V-shaped pattern with a minimum flux 
around the wavelength at rest of 3000 \AA \ is observed  (e.g., Yi et al. 2011, middle panel of Fig. 3), characteristic of a two-component synthesis. The analysis of Yi et al. (2011)
with SDSS-GALEX galaxies showed that 30-40\% of the galaxies have a conspicuous V-shape, 
which is the signal of a young component similar to what was observed in the 
sample of L17 at high redshift: see Fig. \ref{Fig:avfitphotl2}.
The Balmer break is correlated with the (far UV)-r color (Rich et al. 2005, middle panel of Fig. 2), which also indicates that the higher the young component contamination (the lower the (far UV)-r color), the lower the average age (the lower
the Balmer break). 
Therefore, the claim of a pure single stellar population cannot be supported 
in general.

There are other possibilities to explain the UV excesses that
do not include a young component, such as a UV upturn, defined as rising flux with decreasing wavelength from about 2500 \AA \ to 1000 \AA  \ (e.g., Bertola et al. 1982; Yi et al. 2011; Boissier et al. 2018). 
The UV upturn was found in the nearby Universe in quiescent galaxies and has originally been associated with the production of very
hot stars by a young stellar population (O'Connell 1999), but it has also been associated with old stellar systems (Ferguson 1999, Brown 2004) such as old open  (Buzzoni et al. 2012) and globular cluster environments (Perina et al. 2011).
Several theoretical works studied the excess UV emission of various types of stars during their advanced evolutionary phases (Greggio \& Renzini 1990; D'Cruz et al. 1996), although stellar evolution alone cannot provide a full explanation of the UV upturn (Greggio \& Renzini 1990). Explanations were given in terms of extremely horizontal branch stars, in which low-mass helium-burning stars have lost their hydrogen-rich envelope (Brown 2004; Han et al. 2007), or the effect of binarity, with stars losing their hydrogen envelopes during binary interactions (Han et al. 2007; Hern\'andez-P\'erez \& Bruzual 2014). Yi et al. (2011) also considered the UV upturn phenomenon in quiescent galaxies. They showed that 5\% of their sample (within SDSS-GALEX low-z elliptical galaxies in clusters) show a UV upturn and 27\% of them have residual star formation, which means that although the UV upturn may be another effect to take into account, it is not the dominant one in their sample. Similarly, the UV upturn, regardless of its origin, is not the dominant 
contribution in the spectral range aroung 2000-2500\,\AA .
Fig. \ref{Fig:avfitphotl2} shows that a young stellar component fits the average luminosites at rest in the L17 sample, whereas a UV upturn (merely with a linear or almost-linear $\log F(\lambda )$, as has commonly been fit by other authors; e.g., Yi et al. 2011, top panel in Fig. 3; Boissier et al. 2018, top panel in
Fig. 1) does not because the UV luminosity decreases below $\sim $1500 \AA . There is a clear bump in the UV regime, expected in the presence of a young population of $\lesssim 100$ Myr, but not for the UV upturn. 
The models that use an almost linear $\log F(\lambda )$ below 1500 \AA \ might be missing some
deeper effect of the Lyman break, but even so, a functional shape as observed in Fig.
\ref{Fig:avfitphotl2} may not be reproduced in these terms.
Therefore, even if we admit that some galaxies might present
a UV upturn as well, we consider that it is not enough to explain the observed UV features.
However, a fit with two SSPs can even explain the UV upturn in the few galaxies that present that feature with a very young population ($\lesssim $ 25 Myr; L17), therefore we keep the  two-SSP model as a valid way to fit the spectrophotometry of the galaxies and do not consider any further effect of possible old populations with binarity effects, hot horizontal branch, post-AGB, or others (stellar populations with ages $\gtrsim 2$ Gyr
are excluded given the high redshift of our sample).
To summarize,
we do not discard the possibility that there might be some effect from the old populations in UV excess, but we consider that at least some amount of young population is necessary.
Therefore, estimates for the amount of the young contribution represent an approximate
maximum limit.

\begin{figure}[htb]
\vspace{1cm}
\centering
\includegraphics[width=8cm]{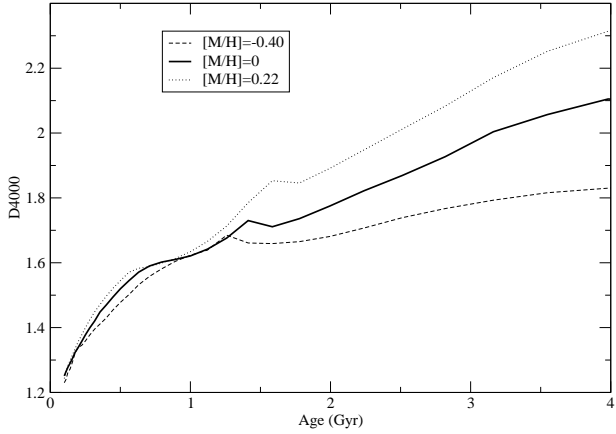}
\vspace{0.5cm}
\caption{Dependence of the Balmer break on the age and metallicity of old, passively
evolving galaxies according to the E-MILES synthesis model.}
\label{Fig:figure1}
\end{figure}

\begin{figure}[htb]
\vspace{1cm}
\centering
\includegraphics[width=8cm]{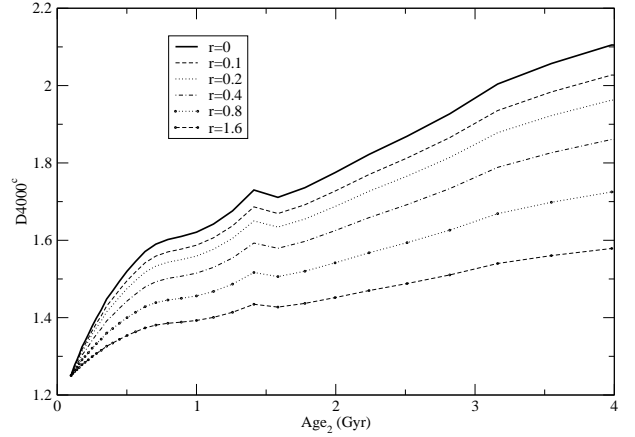}
\vspace{0.5cm}
\caption{Measured Balmer break for two SSPs with age$_1$=0.1 Gyr and age$_2$ and
$r\equiv \frac {F_{\lambda ,{\rm young}}(4\,000\ \AA \ {\rm at\;rest})}{F_{\lambda ,{\rm old}}(4\,000\ \AA \ {\rm at\;rest})}$ according to E-MILES synthesis model with solar metallicity.}
\label{Fig:miles_d4000c}
\end{figure}

The cosmic chronometer method measures the differential age as a function of redshift. 
The Hubble parameter may be calculated as $H(z)=-\dot{z}\,(1+z)^{-1}$, from
which one may infer that (Moresco et al. 2012, 2016) 
\begin{equation}
H(z)=-\frac{1}{1+z}A\frac{dz}{dD_{4000}}\;,
\end{equation}
where the Balmer break $D_{4000}\equiv 
\frac{\overline {F}_\lambda (4050-4100 \AA)}{\overline {F}_\lambda (3900-3950 \AA)}$.
This may be approached as $D_{4000}=A\cdot{\rm age}+B$ with one SSP, $A>0$, $B>0$. This is only a very rough approximation because, as Fig.~\ref{Fig:figure1} shows, the dependence 
of the Balmer break on age is not linear in the young-age regime.  Nonetheless, this expression is commonly used with the cosmic chronometer method (indeed, it is used only for old ages, with $D_{4000}>1.65$ and
with two linear regimes instead of one; Moresco et al. 2012). For instance,
from a fit to Fig.~\ref{Fig:figure1} for solar metallicity, the values of a linear 
fit in the regime between 1 and 4 Gyr would give $A=0.17$ Gyr$^{-1}$, $B=1.46$. 
With a double component of young+old stars with respective ages
age$_1$ and age$_2$ Gyr, however, the combined Balmer break with the two populations is instead
\begin{equation}
D_{c,4000}(z)=\frac{D_{4000}[{\rm age}_2(z)]
+r\,D_{4000}[{\rm age}_1(z)]}{1+r}\;
,\end{equation}
where $r\equiv \frac
{F_{\lambda ,{\rm young}}(4\,000\ \AA \ {\rm at\;rest})}{F_{\lambda ,{\rm old}}(4\,000\ \AA \ {\rm at\;rest})}$, that is, the ratio of luminosities due to the young and old population for the Balmer break. If we use the measured $D_{c,4000}(z)$ without taking into account the contamination of the young population (see Figure \ref{Fig:miles_d4000c}), we obtain a large deviation with
respect to the value of age$_2$ for high values of $r$.
If we assume a linear variation of the young/old population such that $r(z)=\alpha z+\beta $, 
and a constant $age_1$, the measured Hubble parameter $H^*(z)$ is related to the true Hubble parameter $H(z)$ by the
expression 
\begin{equation}
\label{hast}
H^*(z)=-\frac{A}{1+z}\times \frac{1+r}{\frac{dD_{4000}[{\rm age}_2(z)]}{dz}+\alpha D_{4000}[{\rm age}_1]}=
\frac{1+r}{H(z)^{-1}-s}\;,
\end{equation}
where
\[
s= \frac{\alpha D_{4000}[{\rm age}_1](1+z)}{A}\;.
\]

There are other indicators of age, such as the H$_\beta $ Lick index, but this 
was also shown to be affected by the possible presence of ionized gas even in 
the most massive and passive galaxies (Concas et al. 2017).

\subsection{Ratio of luminosities in a high-redshift sample}

L17 fit the spectra of 20 very massive ($\sim 10^{11}$ M$_\odot $) quiescent galaxies at high redshift using two  
such populations with ages of age$_1\lesssim 0.1$ Gyr (young) and age$_2\sim 1.5$ Gyr (old; weighted average, error: 0.5 Gyr). The Balmer break $D_{4000}$ 
assuming solar metallicity would be $<1.25$ and 1.72 for the younger 
and older populations, respectively, using the E-MILES synthesis population model (Vazdekis et al. 2016); see Fig. \ref{Fig:figure1}. The average ratio in this sample is $r=0.71\pm 0.13$, with an
r.m.s. $\sigma _r=0.57$.
Hence, the average Balmer break is $D_{4000}<1.52$, corresponding to an average age 
$<0.5$ Gyr (using the E-MILES model; see Fig. \ref{Fig:figure1}), almost independently of the 
metallicity. This means that using the Balmer break in these cases and assuming a single 
population, the age of the galaxy is underestimated. An average $<0.5$ 
Gyr would be measured instead of the true age of the oldest population,
which is  equal to 1.5 Gyr.

\begin{figure}[htb]
\vspace{1cm}
\centering
\includegraphics[width=8cm]{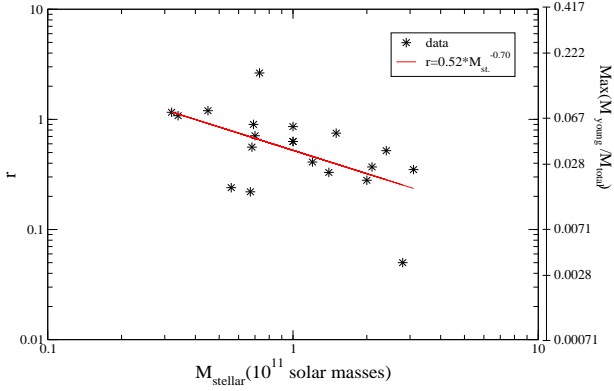}
\vspace{0.5cm}
\caption {Log-log plot of the ratio $r$ vs. an estimated stellar mass of the sample in L17 (neglecting the mass of the young stars and assuming the age of the old population in all of the galaxies to be 1.5 Gyr). In the right side, the equivalent
scale of the maximum ratio of young stellar mass with respect to the total is given, assuming a minimum factor of 14 in the $M_{\rm stellar}/L_{4000\ \AA}$ between old and young population.
The best power-law fit is given by the solid line.}
\label{Fig:r4000vsM2}
\end{figure}

This sample shows some dependence on stellar mass. In Fig. \ref{Fig:r4000vsM2} we show the ratio $r$ versus an estimated stellar mass of the sample in L17. In the calculation of the stellar mass,
we neglected the mass of the young stars, assumed the age of the old population in all of the galaxies to be 1.5 Gyr, and compared it with 
the stellar population synthesis E-MILES model (Vazdekis et al. 2016).
The best power-law fit gives $r=(0.52\pm 0.08)\left(\frac{M_{\rm stellar}}{10^{11} \ M_\odot}\right)^{-(0.70\pm 0.24)}$.
This is a significant mass dependence, which agrees with the results by Caldwell et al. (2003),
who found that low-velocity dispersion galaxies have younger mean ages. 
In cosmic chronometer samples, the samples are constrained to have $M_{\rm stellar}>10^{11}$ M$_\odot $,
which means that $r\lesssim 0.5$ on average.

The impact of a young component ($\lesssim 0.1$ Gyr) with luminosity at 4000 \AA \ equal 
to 50\% of the luminosity at the wavelength of an older population (1.5 Gyr) is plausible 
and not surprising. The mass/luminosity ratio at this wavelength is $>14$ times higher 
for the older population of 1.5 Gyr than for stars with an age $<0.1$ Gyr (using the E-MILES stellar population synthesis model, assuming solar metallicity and a Kroupa universal IMF; Vazdekis 
et al. 2016). Therefore, the young population can indeed significantly contaminate the
spectrum while still representing only $\lesssim 2.4\%$ of the stellar mass. This is compatible with 
the fact that the galaxies are elliptical and dominated in mass by an older population, 
consistent with what is observed in other nearby galaxies (e.g., Yi et al. 2011; Vazdekis 
et al. 2016).  Note that this young component might be associated
with the gas budget remaining after the main star formation episode that took place
1.5 Gyr earlier, which led to the formation of the vast majority of the stellar
populations of the galaxy. In the nearby Universe, this scenario is consistent with a dominant stellar population evolving passively that hosts some residual in situ star formation, as observed in local galaxies. Therefore the young component obtained
in L17 by means of the two-SSP approach should be regarded as a mean luminosity--weighted contribution that is representative of the most recent tail that extends in time after reaching the peak of star formation.

A mean amount of $\lesssim 2.4$\% of the stellar mass in galaxies with total stellar mass of $\sim 10^{11}$ M$_\odot $ means $\sim 3\times 10^9$ M$_\odot $ of young stars, which implies that $\int_0^{t_{\rm max,young}} dt\,[SFR](t)\sim 3\times 10^9$ M$_\odot $, so that the mean star formation rate is $\langle [SFR] \rangle \sim \frac{3\times 10^9\ {\rm M}_\odot }{\Delta T_{\rm young}}$.
The time extent in which the star formation took place, $\Delta T_{\rm young}$, would
be $\sim 200$ Myr if the 100 Myr contribution represented an average of constant [SFR].
However, since the age obtained for the young burst in the two-SSP approach is in reality
a mean luminosity-weighted age, this contribution has to be associated with an even
longer star formation period as the youngest bursts are far more luminous in the UV,
thus biasing our solution toward a younger SSP.
An exponentially decaying star formation model (L17, sect. 6) gives the best fit for [SFR] almost constant ($\tau \rightarrow \infty$) (although this is not as good as the two-SSP solution). Therefore, the box of integration corresponding to the ages of the young population might be extended up to high intermediate ages. 
The youngest end has the greatest effect because of the strong emission in UV, and the oldest component ($\gtrsim 1$ Gyr) because of its overwhelming contribution in mass; the intermediate ages might be present in the galaxies as well, but they provide a 
relatively far lower contribution because of the low mass. 
The 100 Myr like SSP has to be understood as a mean luminosity-weighted
contribution of a young population with an age $\lesssim 1$ Gyr.
This suggest that the mass of the young population is formed during several hundreds of Myr. Therefore, $\Delta T_{\rm young}$ is larger than 200 Myr and the mean $\langle [SFR] \rangle$ is $<15$ M$_\odot $/yr. In our sample, there is no detectable strong star formation, as we do not see strong emission lines (see spectra in L17 for two of these galaxies), as they were photometrically selected to be quiescent galaxies, and the galaxies were required to have no significant emission at [24$\mu $m], expected from dust associated with star formation (Castro-Rodr\'\i guez \& L\'opez-Corredoira 2012). The SFR in the moment in which they were observed must therefore be very low in these galaxies. The possible episodes of star formation of the galaxy have occurred in epochs previous to the observation by several tens or hundreds of Myr, not in the moment in which the galaxies were observed. The SFR may have some fluctuations such that there are periods in which the galaxies have no significant observable star formation activity, and one such low SFR epochs must be the one in which the galaxies are observed.

In the subsample examined by L17 with
an estimated stellar mass higher than $10^{11}$ M$_\odot $, $r=0.47$, with an r.m.s. 
of 0.22 (hence, the error of the mean $r$ is 0.07), if we assume non-variability of this ratio with redshift ($\alpha =0$), we could say 
that for each galaxy in this sample $H^*(z)=(1.47\pm 0.22)H(z)$. That is, if
we used the cosmic chronometer method in this subsample, the value of the Hubble parameter would be overestimated by some unknown factor between 1.25 and 1.69 within 1$\sigma $. We also
note that this $r=0.47$ should represent a maximum limit in this high-redshift sample because possibly all of the UV contribution 
might not necessarily be attributed to
 a young stellar component and there may be 
 other possible contributions associated with the UV upturn phenomenon
(see the discussion of the UV upturn above). Therefore, we should adopt $r\lesssim 0.47$ and
$H^*(z)/H(z)\lesssim (1.47\pm 0.22)$.

We can also perform a double power-law fit with the data using $r(M,z)$, 
for which the result is $r=(0.52\pm 0.08)\left(\frac{M_{\rm stellar}}{10^{11} \ M_\odot}\right)^{-0.37\pm 0.31}\left(\frac{1+z}{3.5}\right)^{-3.3\pm 2.0}$, 
with a slightly significant 
negative $\alpha $; in this case, $s<0$, and $H^*(z)<H(z)$. In either case, this large dispersion of systematic deviations does not allow us to compare $H^*(z)$ measurements with the predictions of various cosmological models.

This may represent a very pessimistic scenario with a very high contamination of young stars. The samples used in the application of cosmic chronometer
methods may be less strongly contaminated, but we do not know the value of $r$ for these samples
because there are no fits using UV photometry in them.

\section{Statistical analysis to fit cosmological models}
\label{.stat}

The comparison of the measured $H(z)$ with the predictions of different cosmological 
models necessarily involves two important aspects: 1) the combination of statistical and 
systematic errors, and 2) the inclusion of other published values of $H(z)$ derived using 
different methods.

\subsection{Combining statistical and systematic errors}
\label{.join}

Most workers in the field, such as Moresco et al. (2012, 2016), simply added statistical 
and systematic errors in quadrature. The justification is based on the view that 
differential dating removes constant systematic uncertainties, so that the residuals 
are uncorrelated. The systematic errors published with the $H(z)$ measurements, however,
include an estimate of stellar metallicity that is measured independently in each redshift 
bin. Although this might provide uncorrelated and independent measurements in principle,
they are nonetheless at least partially correlated.  
At best, one can view these systematic errors
as having some component that one may add in quadrature with statistical errors, 
but not the correlated parts. One way to show that the systematic errors cannot be 
entirely uncorrelated is that the total error resulting from addition in quadrature
is too large compared with the deviations of the data relative to any of the best-fit models (see, e.g., Moresco et al. 2016, Fig. 6; Melia \& Maier 2013; Melia \& Yennapureddy 2018). 
These errors produce a reduced chi-square $\chi ^2_{\rm dof}$ 
that is significantly smaller than one. Moreover, Leaf \& Melia (2017) have shown
that the global errors of $H(z)$ measurements are not Gaussian, therefore they
cannot be purely statistical.

A more appropriate method of combining statistical errors $\sigma _{\rm stat}$ with
systematic errors $\sigma_{\rm syst}$ that are only partially uncorrelated would be
the following (Melia \& Yennapureddy 2018):
\begin{equation}
\sigma^\prime_{\rm stat}=\sqrt{\sigma_{\rm stat}^2+f_s\sigma_{\rm syst}^2}\;,
\end{equation}
\begin{equation}
\sigma^\prime_{\rm syst}=\sqrt{1-f_s}\sigma_{\rm syst}\;,
\end{equation}
where $f_s$ is an unknown constant that represents the degree of uncorrelation in 
the systematic errors. For $f_s=0$, the systematic errors are completely correlated,
whereas for $f_s=1$, the systematic errors are totally random and can be added with the
statistical errors in quadrature. We assume here that $f_s$ does not depend on redshift. 
However, this unknown quantity cannot be calculated directly, although we can fit $f_s$ 
in order to ensure that the optimized fit of the best model has $\chi^2_{\rm dof}=1$.

Now, even if we knew the true systematic error $\sigma _{syst.}'$ and the true
statistical error $\sigma _{stat.}'$, the method for testing cosmological models cannot be 
based solely on summing them quadratically and then minimizing $\chi ^2$, or maximizing
the likelihood function. This is typically done in cosmology, based on a misconception 
in the statistical analysis, but it is not a correct approach. Cosmologists often show 
merit functions with contours representing the confidence regions that constrain the 
optimized parameters, but systematic errors cannot be included in this manner. There is no way to calculate the error bars of a parameter in a model when we have both 
statistical and systematic uncertainties. At best, we can only tell whether a given
model with certain parameter values is compatible with the data, but no 
confidence level in the merit functions can be plotted. In other words, we cannot 
treat systematic errors as if they were uncorrelated and Gaussian.

One possible way to combine the different errors in $H(z)$ is to carry out the fitting
by generating a family of solutions $H_t(z)=H(z)+t\sigma _{syst.}'(z)$, with all 
values in the range $-1\le t\le 1$ and assigning to each $H_t(z)$ a statistical error 
$\sigma _{stat}'(z)$. We here assume that the deviation of $H_t$ with respect 
to the measured $H$ is in the same direction as, and by the same ratio of, the systematic 
error. Of course, this prescription is not unique, but it is a much better approach 
than simply summing the statistical and systematic errors in quadrature. We use this
approach to fit the cosmological models for each $H_t(z)$ using, for instance, the 
minimization of $\chi ^2$, or the maximization of the likelihood function, and we 
choose the value of $t$ corresponding to the smallest of the minima in $\chi ^2$ 
(or the largest of the maxima if we use the likelihood function). The probability 
associated with the smallest of the minima in $\chi ^2$ gives us the maximum 
probability, $P_{\rm max}$, of the model to fit the data.

\subsection{Locally measured values of $H_0$}

Another problem with the use of cosmic chronometers for measuring $H(z)$ is the
inclusion of $H_0\equiv H(z=0)$ derived with other data, such
as Type 1a supernovae (SNe) at
$z\lesssim 0.1$. This should not be done because the calibration used for cosmic 
chronometers is subject to different systematics than those of other methods
(Wei et al. 2017). Furthermore, $H_0$ is an amplitude for all the $H(z)$ measurements using cosmic chronometers, and is not an independent datum. It
should be treated as a free parameter in a fit, while the introduction of 
an a priori value for it produces deviations with respect to the rest of
the global fit based solely on cosmic chronometer data.  Wei et al. (2017 and
references therein) argued that a local measurement of $H_0$ would not be the same as that
derived for the background if the local expansion rate were influenced by 
a Hubble bubble. 

Surprisingly, the value of $H(z)$ in the lowest $z$ bins obtained from
cosmic chronometers is very close to the value $H_0$ obtained by the cosmological
groups using Cosmic Microwave Background Radiation (CMBR), gravitational lensing, 
or SNe. For instance, the value of
$H(z=0.179)=75.0\pm 3.8 {\rm (stat.)}\pm 0.5 {\rm (syst.)}$ km s$^{-1}$ Mpc$^{-1}$ by Moresco et al. (2012) using galaxies of the SDSS survey and the
Bruzual \& Charlot (2003, hereafter BC03) synthesis model.
Assuming standard cosmology with $\Omega _m=0.3$, 
this would imply a calibration of $H_0=68.7\pm 3.5 {\rm (stat.)}\pm 0.5$ (syst.)
km s$^{-1}$ Mpc$^{-1}$, which is very close to the values of
$73.2\pm 1.7$ km s$^{-1}$ Mpc$^{-1}$ from SNe data (Riess et al. 2016) or 
$67.8\pm 0.9$ km s$^{-1}$ Mpc$^{-1}$ from CMBR (of the Planck mission) and lensing data (Planck Collaboration 2016). In addition
to the recognized fact that SNe measurements and CMBR measurements
differ, we may admire here
that the cosmic chronometer value of $H_0$ is also quite accurate and compatible
with other cosmological measurements. 
The fact that they get $H_0^*\approx H_0$ with high accuracy implies that
1) either $r\approx 0\pm 0.05$ and $\alpha =0$ (no evolution of the ratio $r$), that is,
no contamination of a young component ($r=0$), because
for a non-evolving case of constant $r(z)$, $H_0^*=(1+r)H_0$ (from Eq. \ref{hast}); or
2)  $r$ evolves ($\alpha \ne 0$), and then $H_0^*\approx H_0$ can be
obtained with the constraint $\frac{\alpha }{\beta }\sim -\frac{3}{7}$ (assuming
$D_{\rm 4000}\approx 2.0$, $z\approx 0$, $H_0=70$ km\,s$^{-1}$Mpc$^{-1}$=0.070 Gyr$^{-1}$,
$A\approx 0.06$ Gyr$^{-1}$ (Moresco et al. 2012, model BC03, solar metallicity, high $D_{\rm 4000}$)). We know that condition 1) cannot be true, Moresco et al. cannot avoid
within such low values of $r$ the possible contamination of young population using only SDSS data, without UV spectra. We repeat
that Yi et al. (2011) demonstrated that
the contamination of residual young components is also important in SDSS quiescent galaxies at low $z$. The coincidence that we obtain the same value of $H_0$ as other cosmological groups might be due to a coincidence by which
$r$ is significantly larger than 0 and $\frac{\alpha }{\beta }\sim -\frac{3}{7}$, or
to the fact that there are other systematic errors that compensate for the effect
of  a young component ($r\ne 0$).
Therefore, the calibration of this $H_0$ should be taken with a grain of salt.

\begin{table*}[tp]
\caption{Theoretical predictions of $-\dot{z}\,(1+z)^{-1}$ (equivalent to $H(z)$ for
all cases except for static models) for 
the cosmological models studied here.}
\label{Tab:model}
\begin{tabular}{llc}
\hline
\noalign{\smallskip}
Model  & Free cosmol. parameters & $-\dot{z}\,(1+z)^{-1}$  \\
\noalign{\smallskip}
\hline
\noalign{\smallskip}
$\Lambda$CDM, standard $\Omega _m=0.3$ & $H_0$ & 
   $H_0\sqrt{0.3(1+z)^3+0.7}$ \\
$\Lambda$CDM, free $\Omega _m$ & $H_0$, $\Omega_{\rm m}$ & 
   $H_0\sqrt{\Omega _m(1+z)^3+(1-\Omega _m)}$ \\
$w$CDM & $H_0$, $\Omega_{\rm m}$, $w_{\rm de}$ & 
   $H_0\sqrt{\Omega _m(1+z)^3+(1-\Omega _m)(1+z)^{3(1+\omega _{de})}}$ \\Einstein-de Sitter & $H_0$ & $H_0(1+z)^{3/2}$ \\
Friedmann open & $H_0$, $\Omega_{\rm m}$ & $H_0\sqrt{\Omega _m(1+z)^3+(1-\Omega _m)(1+z)^2}$ \\
QSSC & $H_0$, $\Omega_{\rm m}$, $\Omega_\Lambda <0$ &
$H_0\sqrt{(1-\Omega _m-\Omega _\Lambda)(1+z)^4+\Omega _m(1+z)^3+\Omega _\Lambda}$\\
$R_h=ct$ and Milne & $H_0$ & $H_0(1+z)$ \\
Static, linear Hubble law & $H_0$ & $H_0\,(1+z)^{-1}$ \\
Static, tired light & $H_0$ & $H_0$ \\ \hline
\noalign{\smallskip}
\hline
\end{tabular}
\end{table*}

\subsection{Model selection statistics}
\label{.statIC}

The minimum deviation between a model $\mathcal{M}$ and data is obtained when the $\chi ^2$ is a minimum. Nonetheless, in order to compare the 
likelihood of different models with different numbers of free parameters, 
we also need to evaluate the effect of adding free parameters to reduce the 
$\chi ^2$. Unfortunately, there is no universal approach to accurately
estimating the probability of a cosmology fitting the data using a single 
formula for penalizing the less parsimonious models. Nonetheless, we
can calculate a range of probabilities by surveying the results of
several representative methods that punish the excessive use of free
parameters using various criteria. In general, all these approaches
agree that the model most likely to be correct is the one with a
minimum value of
\begin{equation}
\label{chi2mod}
\chi ^2_{\rm mod}\equiv\chi ^2+u\,k\;,
\end{equation}
where $k$ is the number of free parameters, and $u=-\frac{\partial \chi ^2}{\partial k}$ is a constant that depends on the statistical method.

For the most common version of the minimization of the reduced chi-square: 
$\chi ^2_{\rm dof}\equiv \frac{\chi ^2}{N-k}\approx \frac{\chi ^2}{N}\left(1+\frac{k}{N}\right)$, for a large number of data points $N>>k$. 
Hence, $\chi ^2_{\rm mod}=N\chi ^2_{\rm dof}$, $u\approx \frac{\chi ^2}{N}\gtrsim 1$. That is, the minimum value of $u$ is around one for a good fit ($\chi ^2\sim N$). In our case, as we show
below, some cosmological models do not produce good fits ($\chi^2>>N$), but this is for cases without an increase in free cosmological parameters except for $t$ and $H_0$ , so that they are not affected by the
deviations of this approximation.

It is now common in cosmology to compare the evidence for and against 
competing models using the information criteria (IC; for some of the 
earliest applications, see, e.g., Takeuchi 2000; Liddle 2004, 2007; Tan \& 
Biswas 2012; Melia \& Maier 2013). These criteria constitute an enhanced goodness of fit that extends the usual~$\chi^2$ criterion by more fairly 
taking into account the number of free parameters in each model. These 
tools prefer the more parsimonious models, unless the introduction of an
additional parameter provides a substantially better fit to the data.  
This approach reduces the possibility of overfitting, that is
to say, an outcome
in which optimizing with a greater number of free parameters simply 
fits the noise. These criteria yield the relative ranking of two or more competing models, 
along with a numerical assessment of the confidence that each model is the 
best, analogous to likelihoods or posterior probabilities in conventional statistical inference. Unlike the latter, however, the IC methods
can be applied to models that are not `nested,' that is, models that are
not specializations of each other.

One of the oldest IC methods, known as the Akaike information criterion 
(AIC; Akaike 1973; see also Burnham \& Anderson 2002, 2004), provides the 
relative ranks of two or more competing models, and also a numerical 
measure of confidence that each model is the best. These confidences are analogous to likelihoods or posterior probabilities in traditional 
statistical inference. With this method, the quantity
$\chi ^2+2\,k$ is minimized (see, e.g., Melia \& Meier 2013, Eq.~5), so that $u=2$.

Other methods have subsequently been developed based on IC, including
the Kullback information criterion (KIC; Cavanaugh 2004) and the
Bayesian information criterion (BIC; Schwarz 1978). KIC and BIC are 
symmetrized versions of the probability density functions of the 
``true'' model and the model being tested (Cavanaugh 1999), whereas 
AIC does not symmetrize them. For KIC, $u=3$, a value based on a 
derivation in information theory with close ties to statistical 
mechanics (see also Bhansali \& Downham 1977). The corresponding BIC 
has $u=\ln N$, where $N$ is the number of data points. This is 
a misnomer because it is not really based on information theory; it is 
instead an asymptotic ($N\to\infty$) approximation to the outcome of 
a conventional Bayesian inference procedure for deciding between models 
(Schwarz 1978). Tests have shown that the BIC outperforms other such 
criteria used in model selection when the sample is very large (see, 
e.g., Liddle 2004, 2007). In this paper, we have 17 data points, 
for which $\ln N$ is close to $3$.

In a comparison of two or more competing models, $\mathcal{M}_1,\dots,\mathcal{M}_N$, 
the model with the smallest $\chi ^2_{\rm mod}$ is assessed to be the model 
nearest to the `truth,' that is, to the actual correct (although perhaps still) unknown 
model~${\cal M}_*$ that produced the data.  More quantitatively, the models may be 
ranked as follows. If model~$\mathcal{M}_\alpha$ has $\chi ^2_{\rm mod,\alpha}$, the 
non-normalized likelihood that $\mathcal{M}_\alpha$~is closest to the truth is the 
weight $\exp(-\chi ^2_{\rm mod,\alpha }/2)$, with 
\begin{equation}
\label{eq:lastAIC}
{\cal L}(\mathcal{M}_\alpha)=
\frac{\exp(-\chi ^2_{\rm mod,\alpha}/2)}
{\exp(-\chi ^2_{\rm mod,1}/2)+\dots+\exp(-{\rm \chi ^2_{\rm mod,N}}/2)}
\end{equation}
of being closest to the correct model.  In comparing a pair of models 
$\mathcal{M}_1,\mathcal{M}_2$, the difference $\Delta \chi ^2_{\rm mod}\equiv
\chi ^2_{\rm mod,2}-\chi ^2_{\rm mod,1}$ determines the extent to which 
$\mathcal{M}_1$ is favored over~$\mathcal{M}_2$. For $\Delta \chi ^2_{\rm mod}<2$,
the evidence in favor of model $\mathcal{M}_1$ is considered to be weak. The evidence is mildly in favor of $\mathcal{M}_1$ for $2\lesssim \Delta \chi ^2_{\rm mod}
\lesssim 5$, and it is strong for $\Delta \chi ^2_{\rm mod}\gtrsim 5-6$.

To summarize, we minimize $\chi ^2_{\rm mod}$ defined in Eq. (\ref{chi2mod}) 
with a value of $u$ between 1 and 3, and we calculate the corresponding 
range of probabilities using Eq. (\ref{eq:lastAIC}).

\section{Model selection}
\label{.fitting}
Moresco et al. (2012, 2016) have focused their use of cosmic chronometer data
exclusively on the optimization of parameters in the standard model. We can 
broaden the applicability of these observations by comparing different 
cosmological models. We analyze and compare the models 
listed in Table~\ref{Tab:model}, along with their individual predictions
for $H(z)$. Melia \& Yennapureddy (2018) also tried to compare these data with some of these cosmological models, but they restricted their comparison to the $w$CDM/$\Lambda $CDM model
with only one free parameter (including $H_0$), whereas here we leave more unrestricted
possibilities of fitting the model with different cosmological parameters.
We fit a $w/\Lambda $CDM cosmological model with either 1 ($H_0$), 2 ($H_0$, $\Omega_m$) or 3 ($H_0$, $\Omega _m$ and the equation of state for dark energy $w_{\rm de}$) free cosmological parameters. Two settings are valid: either leaving $\Omega_m$, $w_{\rm de}$ as free parameters or taking them fixed within the ``standard'' values to check that this model with these parameters does reproduce the data of cosmic chronometers; and the effect of including a higher
number of free parameters is taken into account with the method explained in \S \ref{.statIC}.
We assume, however, that $\Omega _{\rm total}\equiv \Omega _m+\Omega _\Lambda $ is constant, equal to unity,
and do not let this parameter vary.
We note that for all the 
expanding models, that is, for all models but the last two, $H(z)=-\dot{z}\,(1+z)$. 

\subsection{Data}
We focus on the 17 measurements of $H^*(z)$ taken from the 
literature that have a published separation of the statistical and systematic errors. 
These values are listed in Table~\ref{Tab:data}. We caution, however, that these
systematic errors do not include the progenitor bias, as they are considered negligible
with respect to other uncertainties (Moresco et al. 2012, 2016).

\begin{table}[htb]
\caption{Hubble parameter $H^*(z)$ measured with cosmic chronometers.}
\label{Tab:data}
\begin{tabular}{crrrl}
\hline
\noalign{\smallskip}
$z$  & $\;H^*(z)$ & $\;\sigma_{\rm stat}$ & $\;\sigma_{\rm syst}$ & Reference \\
\noalign{\smallskip}
\hline
\noalign{\smallskip}
0.179 & 75.0 &  3.8 & 0.5 & Moresco et al. (2012)/BC03 \\
0.199 & 75.0 &  4.9 & 0.6 & Moresco et al. (2012)/BC03 \\
0.240 & 79.7 &  2.3 & 1.3 & Gazta\~naga et al. (2009) \\
0.352 & 83.0 &  13.0 & 4.8 & Moresco et al. (2012)/BC03 \\
0.380 & 83.0 &  4.3 & 12.9 & Moresco et al. (2016) \\
0.400 & 77.0 &  2.1 & 10.0 & Moresco et al. (2016) \\
0.420 & 86.4 &  3.3 & 1.7 & Gazta\~naga et al. (2009) \\
0.425 & 87.1 &  2.4 & 11.0 & Moresco et al. (2016) \\
0.450 & 92.8 &  4.5 & 12.1 & Moresco et al. (2016) \\
0.478 & 80.9 &  2.1 &  8.8 & Moresco et al. (2016) \\
0.593 & 104.0 & 11.6 &  4.5 & Moresco et al. (2012)/BC03 \\
0.680 & 92.0 &  6.4 & 4.3 & Moresco et al. (2012)/BC03 \\
0.781 & 105.0 & 9.4 & 6.1 & Moresco et al. (2012)/BC03 \\
0.875 & 125.0 & 15.3 & 6.0 & Moresco et al. (2012)/BC03 \\
1.037 & 154.0 & 13.6 & 14.9 & Moresco et al. (2012)/BC03 \\
1.363 & 160.0 & 22.1 & 24.0 & Moresco (2015, priv. comm.) \\
1.965 & 186.5 & 40.9 & 28.0 & Moresco (2015, priv. comm.)  \\ \hline
\noalign{\smallskip}
\hline
\end{tabular}
\end{table}

\begin{table*}[htb]
\caption{Ranking of minimum probabilities in 
the cosmological models assuming $H(z)=H^*(z)$ [$r=0$] and $f_s=0.160$. 
$H_0$ is given in units of km s$^{-1}$ Mpc$^{-1}$. The probability ranges give their minimum and
maximum value within the three variants of the $w$CDM/$\Lambda $CDM model, between 1 and 3 free cosmological parameters, with $u$ between 1 and 3 and $t$ between -1 and 1.}
\label{Tab:fits}
\begin{tabular}{llcc}
\hline
\noalign{\smallskip}
Model  & Optimized free parameters & $\chi ^2[-1\le t\le 1]$ & Probability range \\
\noalign{\smallskip}
\hline
\\
\noalign{\smallskip}
$\Lambda$CDM, standard $\Omega _m=0.3$ & $t=0.25$, $H_0=69.3\pm 1.6$ & 15.00-29.82 & 0.417-0.989 \\
$\Lambda$CDM, free $\Omega _m$ & $t=0.32$, $H_0=68.5^{+4.0}_{-5.0}$, $\Omega_{\rm m}=0.33^{+0.19}_{-0.13}$ & 
14.84-31.38 & 0.295-0.953 \\
$w$CDM & $t=0.16$, $H_0=79.0^{+56.3}_{-17.0}$, $\Omega_{\rm m}=0.31^{+0.22}_{-0.24}$, $w_{\rm de}=-1.9^{+1.6}_{-2.8}$ & 13.42-26.29 & 0.0514-0.994 \\
$R_{\rm h}=ct$ and Milne & $t=0.49$, $H_0=62.6\pm 1.4$ & 18.50-39.00 & $4.7\times 10^{-3}$-0.733 \\
Friedmann open & $t=0.53$, $H_0=62.2^{+1.7}_{-3.3}$, $\Omega_{\rm m}=0.04^{+0.33}_{-0.04}$ & 18.43-40.09 & $1.3\times 10^{-3}$-0.331  \\
QSSC & $t=0.77$, $H_0=57.7^{+2.1}_{-3.8}$, $\Omega_{\rm m}=1.32^{+0.42}_{-0.18}$, $\Omega_{\Lambda}=0^{+0}_{-0.30}$  & 24.30-52.66 & $5.4\times 10^{-7}$-0.0235 \\
Einstein-de Sitter & $t=1.00$, $H_0=53.4\pm 1.2$ & 38.04-78.93 & $1.0\times 10^{-11}$-$1.4\times 10^{-4}$ \\
Static, tired light & $t=-1.00$, $H_0=79.0\pm 1.9$ & 58.51-106.19 & $9.4\times 10^{-20}$-$5.2\times 10^{-6}$  \\
Static, linear Hubble law & $t=-1.00$, $H_0=104.1\pm 2.6$ & 161.51-288.17 & $<2.2\times 10^{-28}$ 
\\\\ \hline
\noalign{\smallskip}
\hline
\end{tabular}
\end{table*}

\subsection{Assuming that the measurements of $H(z)$ are correct}
\begin{figure*}[htb]
\vspace{1cm}
\centering
\includegraphics[width=14cm]{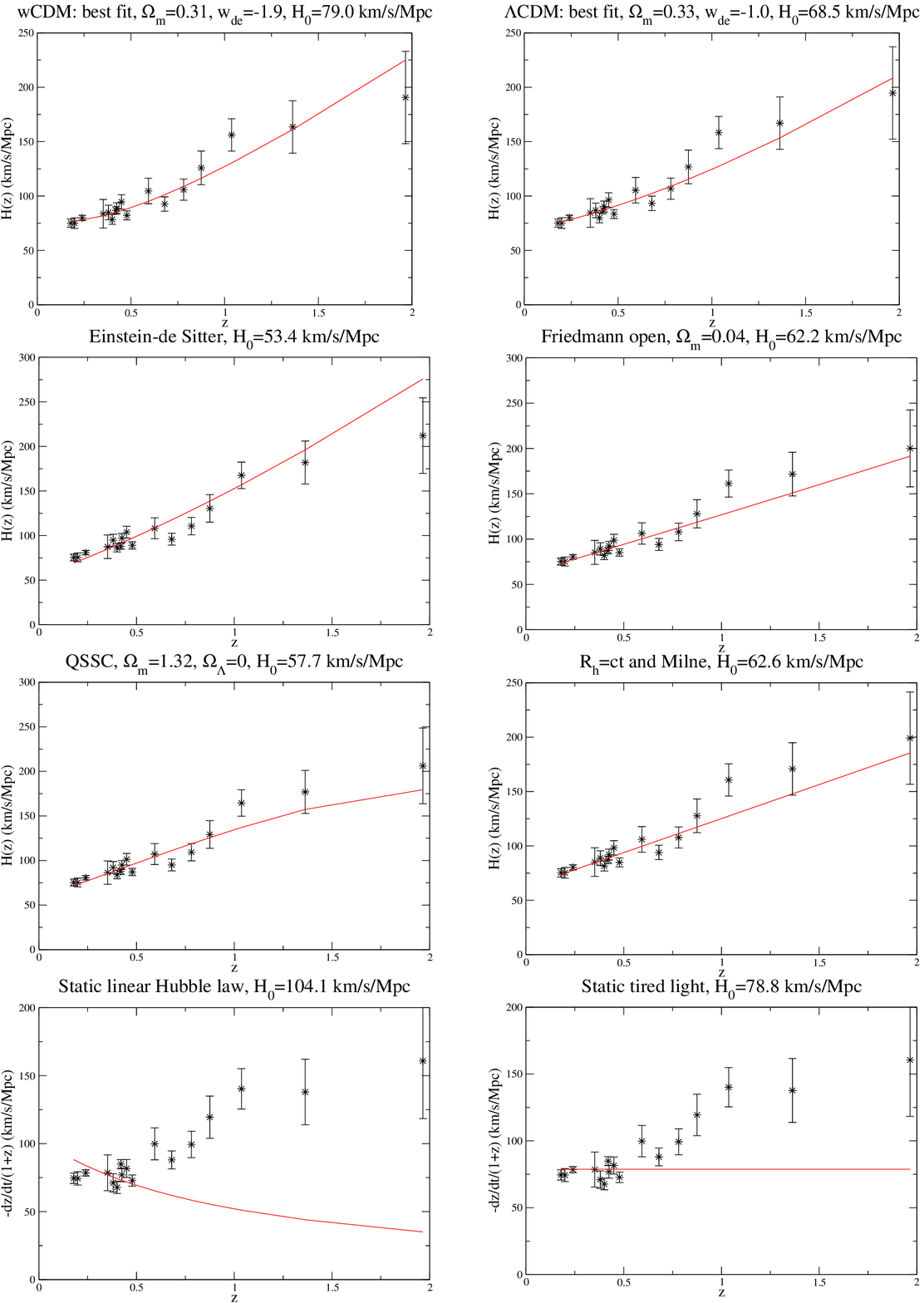}
\caption{Best fits corresponding to the optimized value of $t$, assuming $H_t(z)=H_t^*(z)$ [$r=0$] for different cosmological models. Error bars stand for
$\sigma_{\rm stat}'$. The data are somewhat different in each case because for each 
cosmological model, a different value of $t$ is assigned within the allowed
systematic error.}
\vspace{1cm}
\label{Fig:figure2}
\end{figure*}

We first assume that $H^*(z)=H(z)$ in Equation~(\ref{hast}), 
that is, that there is no contamination by a young stellar component 
and thus $r=s=0$. We follow the procedure outlined in \S~\ref{.join}. 
To estimate the fraction $f_s$, we first fit the data with $\Lambda$CDM and
a fixed standard value of $\Omega_m=0.3$ to obtain a $\chi ^2_{dof}=1$ (the number of free parameters here is two: $t$ and $H_0$, so $N_{dof}=15$).
We obtain $f_s=0.160$.
The ranking of the various cosmological models was calculated
relative to this optimization. The results of the fits for all the cosmologies 
are shown in Fig. \ref{Fig:figure2} and listed in Table~\ref{Tab:fits}.

The optimized free parameters correspond to the best fits for the value of $t$ 
that minimizes $\chi ^2(t)$. To calculate the $1\sigma$ errors, we used the standard
$\chi^2$ test (Avni 1976) with the number of free parameters equal to the number the 
free cosmological parameters (1-3) plus one (i.e., $t$). The probabilities quoted in 
this table are calculated according to Equation~(\ref{eq:lastAIC}), and their range 
includes the minimum and maximum values corresponding to $t$ between -1 and 1, $u$ 
between 1 and 3, for the three variants of the $w$CDM/$\Lambda $CDM model, 
and the assumption that only one of seven possibilities (there are 
seven possibilities in nine different cosmological models 
because Milne and $R_h=ct$ have the same prediction, and 
wCDM and $\Lambda $CDM do not exclude each other) may be correct.
The favored model is the model $w$CDM/$\Lambda $CDM, except in the
case of $u=3$ and three free cosmological parameters for this model, for which $R_h=ct$ or Milne obtain a slightly higher probability for some values of $t$. Previous claims of a better fit for $R_h=ct$ model are based either on using only tests with the largest 
penalization on the large number of free parameters (Melia \& Maier 2013; Melia \& McClintock 2015; Wei et al. 2017) or using $w$CDM/$\Lambda $CDM allowing one free parameter at most, including $H_0$ (Melia \& Yennapureddy 2018), which
restricts the possibilities for exploring the full space of parameters in the standard cosmology.
Here, we see that a global analysis with all possible model selection statistics and without restrictions a priori of some cosmological parameters favors the standard model over the $R_h=ct$ model.
There is also some support for the Friedmann open model. On the other 
hand, the quasi-steady state cosmology (QSSC) is rejected at $>95$\% CL in every
case. Other models, such as the Einstein-de Sitter model, the static tired light model, and 
static linear Hubble law cosmologies, are rejected very strongly. At this point, 
we have not yet taken into account the potentially large systematic errors that are due 
to contamination from young stars, as explained in \S~\ref{.young}. We examine this effect next.

\subsection{Effect on $H(z)$ from the contamination of young stars}
We now assume that the true Hubble parameter is related to the measured values
given in Table \ref{Tab:data} through Eq.~(\ref{hast}), leaving the ratio $r$ 
(although constant in $z$) as a free parameter. The additional systematic error 
corresponding to this correction is
$\Delta H(z)=H(z)-H^*(z)=-r(1+r)^{-1}H^*(z)$.

\begin{table*}[htb]
\tiny
\caption{Value of $f_s$ to obtain $\chi ^2_{dof}=1$ in the model 
$\Lambda$CDM, standard $\Omega _m=0.3$ and probabilities assuming $r\ne 0$. The range gives their minimum and maximum value within the three variants of the $w$CDM/$\Lambda $CDM model, between 1 and 3 free cosmological parameters, with $u$ between 1 and 3 and $t$ between -1 and 1.}
\label{Tab:rne0}
\begin{tabular}{lcccccc}
\hline
\noalign{\smallskip}
  & $r=0.1$ & $r=0.2$ & $r=0.4$ & $r=0.8$ & $r=1.6$  \\ \hline
$f_s$ & 0.136 & 0.113 & 0.075 & 0.034 & 0.0033 &  \\ \hline     
\noalign{\smallskip} 
Prob. [$\Lambda$CDM, standard $\Omega _m=0.3$] &  0.360-0.996 & 0.265-0.9986 & 0.182-0.99983 & 0.105-0.9999952
& 0.0578-1.000000 \\
Prob. [$\Lambda$CDM, free $\Omega _m$ &  0.297-0.956 & 0.269-0.968 & 0.232-0.964 & 0.157-0.957 &
0.0849-0.779 \\
Prob. [$w$CDM] &  0.0549-0.996 & 0.0503-0.9983 & 0.0427- 0.99909 & 0.0331-0.9956 
& 0.0212-0.464\\
Prob. [$R_{\rm h}=ct$] or Prob.[Milne] &  $2.4\times 10^{-3}$-0.753 & $1.3\times 10^{-3}$-0.691 &
$1.7\times 10^{-4}$-0.695 & $4.8\times 10^{-6}$-0.733 & $3.9\times 10^{-8}$-0.899 \\
Prob. [Friedmann open] &  $7.1\times 10^{-4}$-0.284 & $3.6\times 10^{-5}$-0.372 & $7.9\times 10^{-7}$-0.384
& $2.0\times 10^{-9}$-0.350 & $<0.309$  \\
Prob. [QSSC] &  $2.7\times 10^{-8}$-0.0447 & $2.5\times 10^{-9}$-0.0593 & $9.4\times 10^{-12}$-0.104 & 
$1.1\times 10^{-15}$-0.168  & $<0.181$ \\
Prob. [Einstein-de Sitter] & $1.4\times 10^{-13}$-$1.1\times 10^{-3}$ & 
$2.8\times 10^{-15}$-$4.7\times 10^{-3}$ & $1.1\times 10^{-18}$-0.0343 & $<0.217$ & $<0.526$ \\
Prob. [Static, tired light] &  $1.4\times 10^{-17}$-$2.8\times 10^{-7}$ & $5.9\times 10^{-16}$-$2.9\times 10^{-8}$  & $1.4\times 10^{-13}$-$9.8\times 10^{-10}$ & $5.0\times 10^{-16}$-$2.6\times 10^{-10}$ & $<1.3\times 10^{-7}$\\
Prob. [Static, linear Hubble law] & 0 & 0 & 0 & $<9.2\times 10^{-29}$ & $<1.3\times 10^{-19}$
\\\\ \hline
\noalign{\smallskip}
\hline
\end{tabular}
\end{table*}

\begin{figure*}[htb]
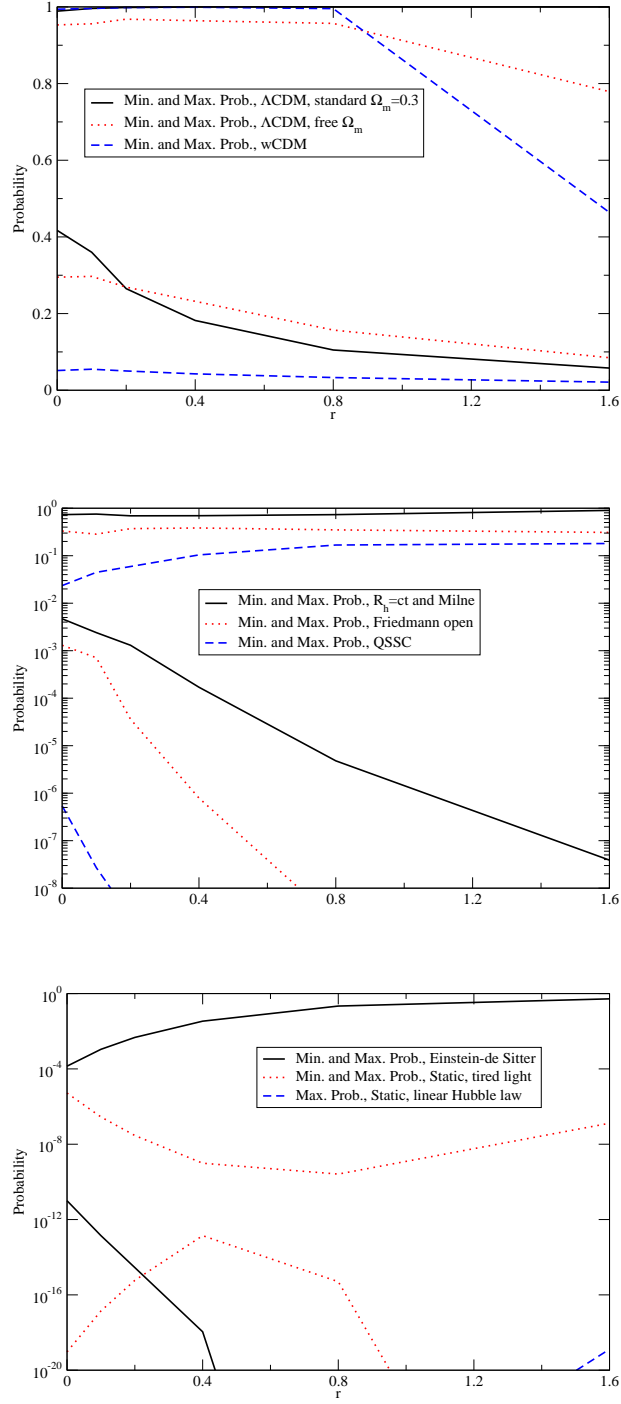

\vspace{1cm}
\centering
\includegraphics[width=8cm]{prob123.eps}\\
\vspace{1cm}
\includegraphics[width=8cm]{prob456.eps}\\
\vspace{1cm}
\includegraphics[width=8cm]{prob789.eps}\\
\caption{Range of probabilities for the different models as a function of $r$, assuming
no evolution ($\alpha =0$).}
\vspace{1cm}
\label{Fig:prob}
\end{figure*}

\begin{figure*}[htb]
\vspace{1cm}
\centering
\includegraphics[width=14cm]{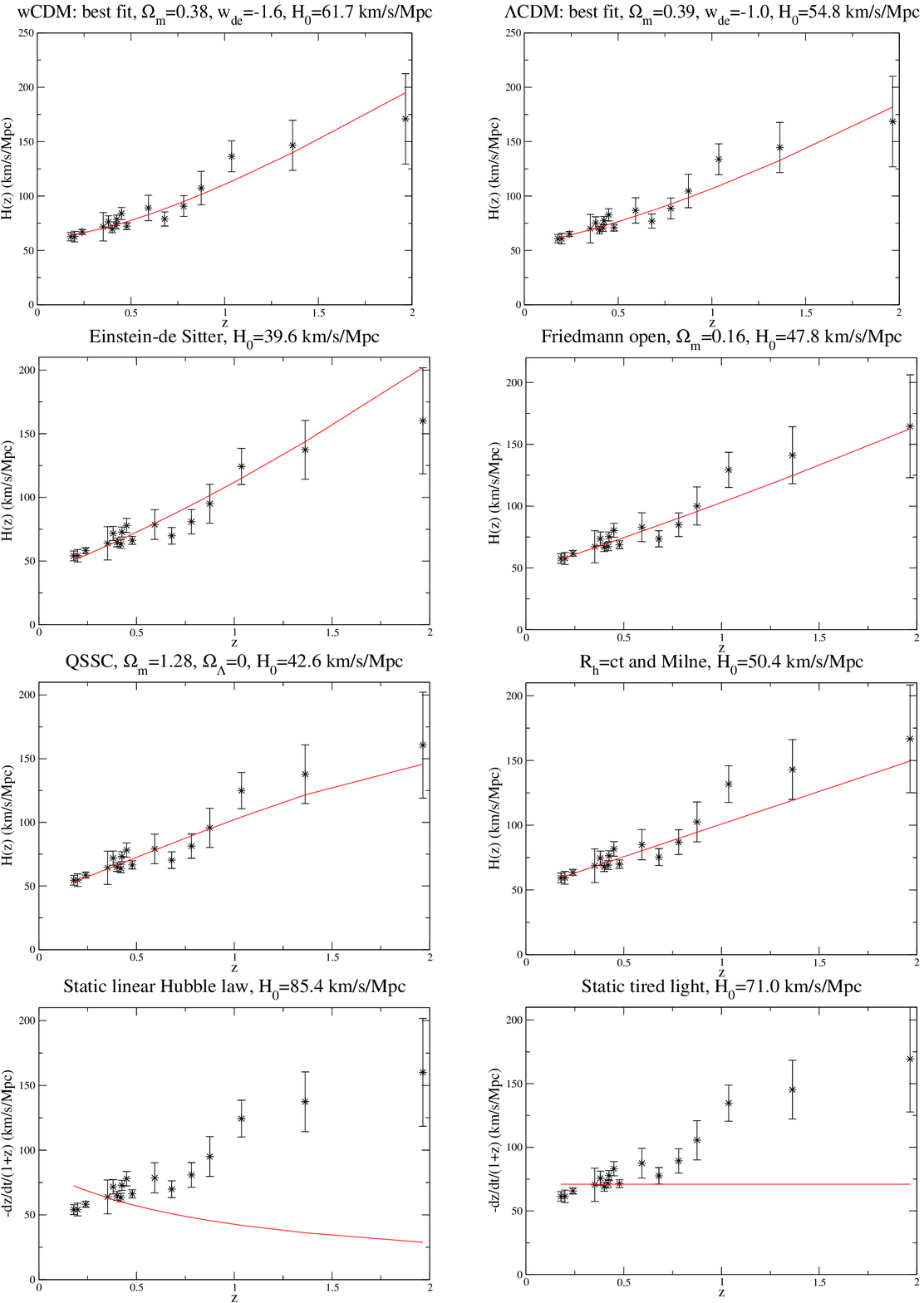}
\caption{Best fits corresponding to the optimized value of $t$, assuming $r=0.4$, $\alpha =0$ for different cosmological models. Error bars stand for
$\sigma_{\rm stat}'$. The data are somewhat different in each case because for each 
cosmological model, a different value of $t$ is assigned within the allowed
systematic error.}
\vspace{1cm}
\label{Fig:fit_cosm_chr04}
\end{figure*}

In Table \ref{Tab:rne0} and Fig. \ref{Fig:prob} we quote the range of probabilities for $r=0.1,0.2,0.4,0.8$ and $1.6$, 
based on the same criterion for the calculation of $f_s$, that
is, using as reference the best 
fit for the standard model without free parameters.
In Fig. \ref{Fig:fit_cosm_chr04} we plot the best fits for the case of $r=0.4$.
The Einstein-de Sitter and 
QSSC models pass the test when we include this systematic error with an average value $r=0.47\pm 0.07$ obtained in the analysis of L17 subsample
with masses higher than $10^{11}$ M$_\odot $. However, the two static models 
are rejected even for values as high as $r=1.6$.

Although the cosmic chronometer method is not reliable enough
to extract accurate cosmological information, we can therefore nonetheless use it to reject
a static Universe at a very high confidence level. When the
systematic error due to possible contamination by a young population
is taken into account, however, the remaining models can pass the test. 

\section{Further discussion and conclusion}
\label{.conclusion}

The cosmic chronometer method uses the age difference between a passively evolving 
population of galaxies at different redshifts to calculate the expansion rate of the 
Universe. This is an elegant way to derive cosmological information, provided that 
the age measurements reflect the time passed since the galaxy was formed. The proposers 
of this method use the relationship between the amplitude of the Balmer break at 
4000 \AA \ at rest and the age of the galaxy, including the possible effects of 
degeneracy with the metallicity, but what they measure is some kind of 
``average'' age of the populations of the galaxy. The fact that there is a 
component of young stars in quiescent galaxies strongly contaminates the luminosity at 
the Balmer break wavelength, producing much lower values than the age of the oldest 
population in that galaxy.

As an example of this contamination, we have
examined the sample of quiescent galaxies at $z>2$ with $M_{\rm stellar}>10^{11}$ M$_\odot $ in L17, which yield an average contribution to the luminosity at 4000 \AA \ from the young component of about 50\% of that due to the oldest population, that is, 33\% of the total luminosity, which is far from negligible. 
This should be considered as an upper limit, 
as there might be other contributions from older stellar populations that might
be associated with the UV upturn phenomenon, although our spectral
energy distribution fitting points to a dominant
contribution from the young component in this spectral range, most likely coming from
the most recent tail of the main star formation event that took place $\sim $1 Gyr earlier.
This ratio is significantly dependent on stellar mass, proportional to $M^{-0.7}$; we will explore this
important dependence in future works in more detail. The degree of contamination may vary in other samples of quiescent galaxies, but it is expected to be significant in all of them. 

Consequently, measurements of the absolute value of the age, or the differential age between different redshifts, are incorrect, making the 
cosmic chronometer method very inaccurate.
Even so, the predictions of $\dot{z}$ by the various cosmological models are very
different, and some exotic scenarios, such as the static Universe models, can be 
rejected.

\begin{acknowledgements}
Thanks are given to F. Melia for his comments and suggestions to improve this paper, to
J. Betancort-Rijo for some discussions on statistical techniques, to M. Moresco for clarifications of his papers and for providing the systematic errors of the points of Moresco (2015), to the anonymous referee for helpful comments, 
and to Astrid Peter (language editor of A\&A) for the revision of this paper.
MLC acknowledges support from grant AYA2015-66506-P 
from the Spanish Ministry of Economy and Competitiveness (MINECO).
AV acknowledges support from grant AYA2016-77237-C3-1-P
from the Spanish Ministry of Economy and Competitiveness (MINECO).

\end{acknowledgements}

\appendix

\end{document}